\documentclass{iau} 
\usepackage{graphicx}         
\usepackage{pifont}
\usepackage{hyperref}
\usepackage{units}
\usepackage{xcolor}

\definecolor{blue}{rgb}{0.00, 0.00, 1.00}
\definecolor{red}{rgb}{0.86, 0.08, 0.24}
\definecolor{orange}{rgb}{1.00, 0.55, 0.00}
\definecolor{darkblue}{rgb}{0.00, 0.00, 0.55}
\definecolor{green}{rgb}{0.00, 0.39, 0.00}
\definecolor{pink}{rgb}{1.000000,0.078431,0.576471}

\newcommand\arcsec{\mbox{$^{\prime\prime}$}}%
\title[Coordinated observations between China and Europe] 
{Coordinated observations between China and Europe to follow active region 12709}

\author[Gonz\'alez Manrique et al.]   
{S. J. Gonz\'alez Manrique$^1$,
 C. Kuckein$^2$, 
 P. G{\"o}m{\"o}ry$^1$,
 S. Yuan$^3$, \\
 Z. Xu$^3$,
 J. Ryb\'ak$^1$,
 H. Balthasar$^2$,
 \and P. Schwartz$^1$}

\affiliation{$^1$Astronomical Institute, Slovak Academy of Sciences, \\
05960 Tatransk\'{a} Lomnica, Slovak Republic \\ email: {\tt smanrique@ta3.sk} \\[\affilskip]
$^2$Leibniz-Institut f{\"u}r Astrophysik Potsdam (AIP), \\
An der Sternwarte 16, 14482 Potsdam, Germany \\[\affilskip]
$^3$Yunnan Observatories, Chinese Academy of Sciences, \\
Kunming, 650011, China }

\pubyear{2019}
\volume{354}  
\jname{Solar and Stellar Magnetic Fields: Origins and Manifestation}
\editors{Kosovichev, Strassmeier \& Jardine, eds.}
\begin{document}

\maketitle

\begin{abstract}
We present the first images of a coordinated campaign to follow
active region NOAA 12709 on 2018 May 13 as part of a joint effort between 
three observatories (China-Europe). The active region was close to disk center 
and enclosed a small pore, a tight polarity inversion line and a filament
in the chromosphere. The active region was observed with the 1.5-meter GREGOR solar telescope 
on Tenerife (Spain) with spectropolarimetry using GRIS in 
the He~{\sc i} 10830 \AA\ spectral range and with HiFI using two broad-band filter channels.
In addition, the Lomnicky Stit Observatory (LSO, Slovakia) recorded the same 
active region with the new Solar Chromospheric Detector (SCD) in spectroscopic 
mode at H$\alpha$ 6562 \AA. The third ground-based telescope was located at the Fuxian Solar 
Observatory (China), where the active region was observed with the 1-meter New Vacuum Solar 
Telescope (NVST), using the Multi-Channel High Resolution Imaging System at H$\alpha$ 6562 \AA. 
Overlapping images of the active region from all three telescopes will be shown as well
as preliminary Doppler line-of-sight (LOS)
velocities. The potential of such observations are discussed.
\keywords{Sun: activity, Sun: chromosphere, Sun: filaments, Sun: magnetic fields}
\end{abstract}

\firstsection 
              
\section{Introduction}
Solar phenomena extend over many layers of the atmosphere. Frequently magnetic fields are
rooted in the photosphere and expand with height and twist, forming helical configurations 
such as filaments. In order to follow the field lines across the solar atmosphere,
multiwavelength observations are crucial. However, current ground-based telescopes 
such as GREGOR only offer up to three simultaneous instruments which cover different
wavelength ranges. To better understand the formation and evolution of solar phenomena 
we need more co-temporal observations with many wavelengths. This is also the main 
objective of future high-resolution telescopes such as DKIST 
(\cite[Tritschler \etal\ 2015]{Tritschler2015}) and EST (\cite[Collados \etal\ 
2013]{Collados2013}). In the meantime, efforts are being done to coordinate high-resolution
ground-based telescopes to study our Sun. This work shows the viability of coordinated
observations between telescopes in Europe and China, as well as the potential when
combining these observations.

\section{Observations}
Active region (AR) NOAA 12709 was observed on 2018 May 13 very close to disk center at 
heliographic coordinates (S6$^{\circ}$, W3$^{\circ}$). The AR was of interest because it enclosed 
several solar phenomena. In the center of the AR, a narrow polarity inversion line separated an extensive area 
of opposite polarities. Furthermore, a decaying pore was present at the border of the AR and an 
intermediate filament was seen in the chromosphere.

\begin{figure}[t]
\begin{center}
 \includegraphics[width=0.9\textwidth]{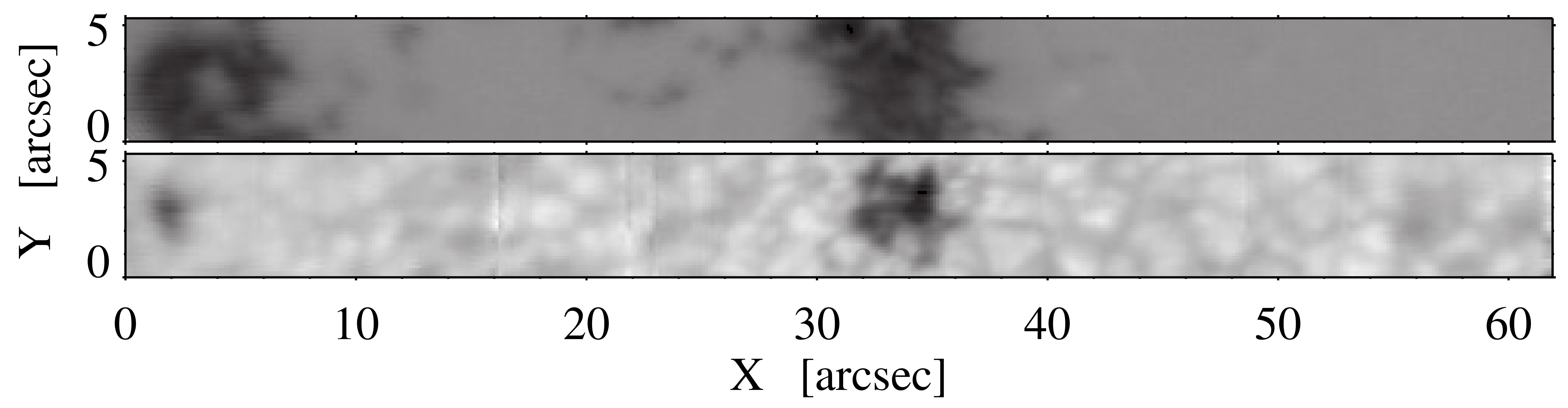} 
 \caption{Continuum intensity (\textit{bottom}) and Stokes $V$ 
          (\textit{top}) slit-reconstructed image of AR
          NOAA 12709 recorded with GRIS between 08:53:13~UT and 08:57:02~UT
          on 2018 May 13.}
   \label{fig2}
\end{center}
\end{figure}

Four instruments, which were located at three different telescopes around the world, were recording 
simultaneously this active region: (1) The GREGOR Infrared Spectrograph (GRIS, 
\cite[Collados \etal\ 2012]{Collados2012}) and (2) the High-resolution Fast Imager (HiFI, 
\cite[Kuckein \etal\ 2017]{Kuckein2017}, \cite[Denker \etal\ 2018]{Denker2018}) 
attached to the 1.5-meter GREGOR solar telescope on Tenerife, Spain (\cite[Schmidt \etal\ 2012]{Schmidt2012}); (3) The new Solar Chromospheric Detector (SCD, \cite[Kucera \etal\ 2015]{Kucera2015})
based on the Lomnicky Stit Observatory (LSO), Slovakia; and (4) the Multi-Channel 
High Resolution Imaging System (\cite[Xu \etal\ 2013]{Xu2013}) placed at the 
the 1-meter New Vacuum Solar Telescope, China (NVST, \cite[Liu \etal\ 2014]{Liu2014}). 

We observed with HiFI anf GRIS around two hours and twenty minutes the same region 
(starting at 08:08~UT), mainly focusing on the main pore but also observing the polarity 
inversion line. The SCD science data were acquired in the time intervals 
05:57\,--\,06:28, 06:43\,--\,07:19, 08:52\,--\,09:27, and 09:53\,--\,10:10~UT. We observed
with a good afternoon seeing at the Fuxian lake around 1.5 hours with the H$\alpha$ filter band  
on 08:48\,--\,10:22~UT while we only observed around 20 minutes with the TiO starting at 9:38~UT.

All data were dark and flat-field corrected, as well as polarimetrically calibrated following standard 
procedures. HiFI and NVST data were restored using the speckle code from
\cite[W{\"o}ger \& von der L{\"u}he (2008)]{Woger2008} and
\cite[Liu \etal\ (1998)]{Liu1998}, respectively.

\section{Combining different telescopes}
\begin{figure}[t]
\begin{center}
 \includegraphics[width=\textwidth]{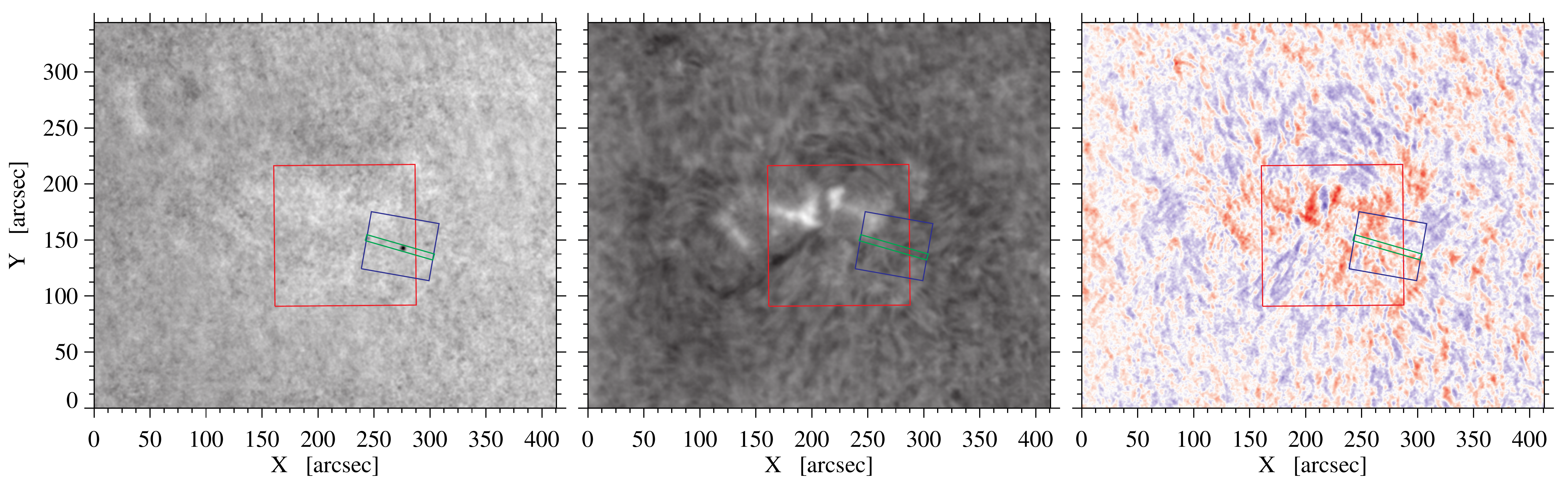} 
  \caption{Overview of AR NOAA 12709 obtained with the SCD
           at 08:53:09~UT on 2018 May 13. Quasi-continuum intensity in the far 
           H$\alpha$ blue line wing (\textit{left}), H$\alpha$ line core 
           intensity (\textit{middle}), and Doppler shifts 
           clipped between $\pm$2~km~s$^{-1}$ (\textit{right}). The red and blue rectangles 
           represent FOVs of NVST and HiFI instrument at GREGOR, respectively. 
           The green rectangle represents the FOV covered by GRIS.}\label{fig1}
\end{center}
\end{figure}

\begin{figure}[t]
\begin{center}
 \includegraphics[width=\textwidth]{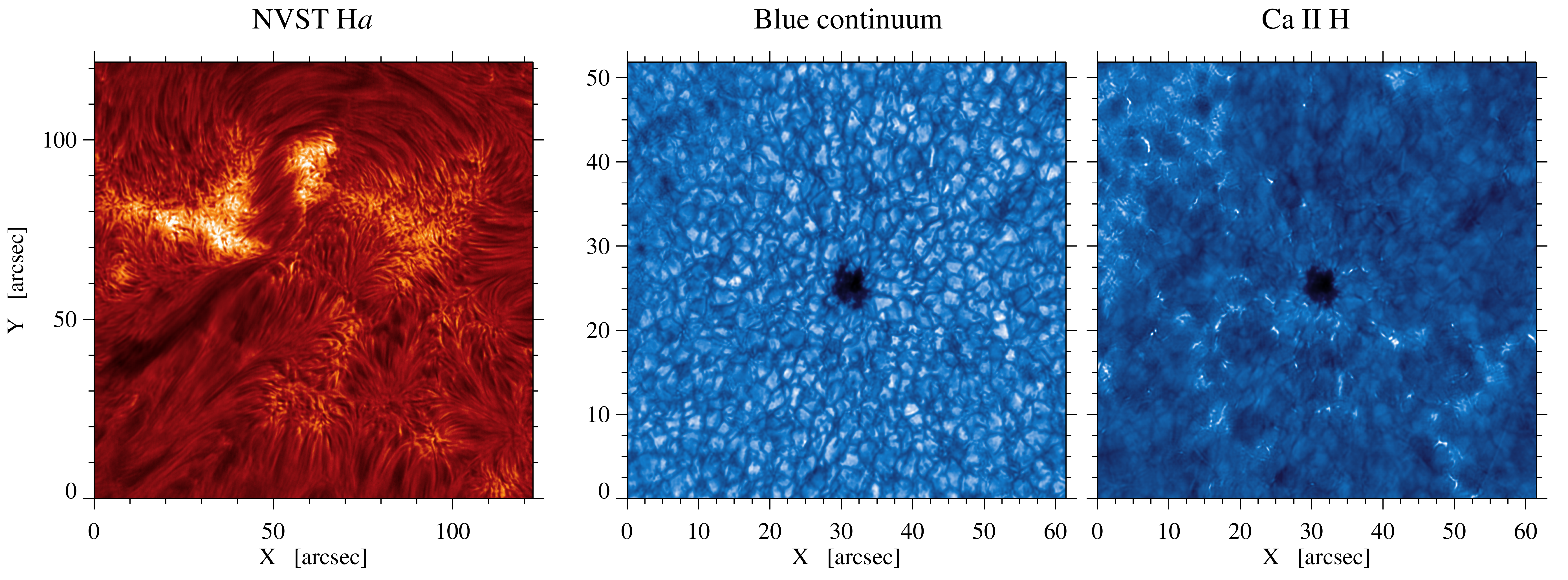} 
  \caption{Overview of the active region NOAA 12709 obtained on 2018 May 13. 
           Restored H$\alpha$ 6562.8 \AA\ image recorded (08:53:05~UT) with 
           the Multi-Channel High Resolution Imaging System placed 
           on the NVST (\textit{left}), and HiFI Speckle-reconstructed 
           (08:53:08~UT) blue continuum 4505 \AA\ (\textit{middle}) 
           and Ca~{\sc ii}~H 3968 \AA\ (\textit{right}). NVST and HiFI are not
           showing the same FOV. }\label{fig3}
\end{center}
\end{figure}

One big challenge is the combination of data acquired with different instruments. While SCD, NVST, and HiFI are imaging instruments,
GRIS is a slit spectrograph. Hence, GRIS only provides slit-reconstructed 2D images (Fig.~\ref{fig2}). Furthermore, all instruments have a different 
image scale. While HiFI provides very high spatial resolution images in the blue wavelength range (about 0.025\arcsec\,pixel$^{-1}$),
a lower spatial sampling of 0.136\arcsec\,pixel$^{-1}$ or 0.340\arcsec\,pixel$^{-1}$ is recorded by NVST and SCD in H$\alpha$, 
respectively. Yet, having different image scales can be beneficial because they produce different sizes of the FOV. SCD provides the largest
FOV with about 410\arcsec$\times$335\arcsec\ (Fig.~\ref{fig1}) and therefore it is an excellent instrument to cover context information surrounding the largely spread AR. 
The images from the 1-meter NVST provide high resolution images of the chromosphere and 
still cover a large FOV of about 126\arcsec$\times$126\arcsec\ (left panel of 
Fig.~\ref{fig3}).

\section{Potential of coordinated multiwavelength observations}
For the analysis of the central pore in Fig. \ref{fig3}, the combination of the present instruments, together with space telescopes, 
have the potential to provide: (1) the vector magnetic field and LOS velocities in the photosphere and chromosphere by applying spectral line inversion tools to the 
four Stokes parameters (GRIS data set); (2) tracking of horizontal
flows using the high-resolution data of HiFI and NVST; (3) Doppler velocities of the chromosphere in the whole AR
using SCD, to study the connectivity of the pore to the entire AR;
(4) context data from the Solar Dynamics Observatory 
(SDO, \cite[Pesnell \etal\ 2014]{Pesnell}) 
will be used to follow the dynamics of the
AR at different coronal layers of the atmosphere. Additionaly, the
Helioseismic and Magnetic  Imager  
(HMI, \cite[Scherrer \etal\ 2012]{Scherrer2012};
\cite[Schou \etal\ 2012]{Schou2012}) provides the possibility to 
investigate the evolution of the vector magnetic field of the whole AR by
carrying out magnetic field extrapolations.

\begin{acknowledgements}
\noindent The 1.5-meter GREGOR solar telescope was built by a German consortium under the 
leadership of the Leibniz-Institut f\"ur Sonnenphysik in Freiburg (KIS) with the
Leibniz-Institut f\"ur Astrophysik Potsdam (AIP), 
the Institut f\"ur Astrophysik G\"ottingen (IAG), the Max-Planck-Institut f\"ur
Sonnensystemforschung in G\"ottingen (MPS), and the Instituto de Astrof\'isica 
de Canarias (IAC), and with contributions by the Astronomical 
Institute of the Academy of Sciences of the Czech Republic (ASCR).
This work is part of a collaboration between the AISAS and AIP supported by the German DAAD, 
with funds from the German Federal Ministry 
of Education \& Research and Slovak Academy of Science, under 
project No. 57449420. SJGM, PG, and PS acknowledge the support 
of the project VEGA 2/0004/16. SJGM also is grateful for 
the support of the Stefan Schwarz grant of the  Slovak Academy of Science.
\end{acknowledgements}

\end{document}